\documentclass[12pt]{article}	
\pdfoutput=1 
\usepackage{cite}
\usepackage{color} 
\usepackage{amsmath}
\usepackage{amsfonts}
\usepackage{amssymb}
\usepackage{array,booktabs}
\usepackage{float}
\usepackage{caption}
\usepackage{subcaption}
\usepackage{graphicx}
\usepackage{slashed}            
\usepackage{bbm}                
\usepackage{xspace}				
\usepackage{collref}				
\usepackage{framed}				
\usepackage{placeins}
\usepackage{jheppub}							
\usepackage{mathrsfs}
\usepackage{hyperref}
\usepackage{cancel}
\usepackage[normalem]{ulem}

\newcommand{\beq}{\begin{equation}}
\newcommand{\eeq}{\end{equation}}
\newcommand{\beqa}{\begin{equation}\begin{aligned}}
\newcommand{\eeqa}{\end{aligned}\end{equation}}
\newcommand{\lsim}{\lesssim}
\newcommand{\gsim}{\gtrsim}

\begin{document}
\title{Reconciling Large- and Small-Scale Structure In Twin Higgs Models}

\author{Valentina Prilepina,$^{1}$, Yuhsin Tsai$^{2}$}
\affiliation{$^1$Physics Department, University of California, Davis Davis, CA 95616\\
$^{2}$Maryland Center for Fundamental Physics, Department of Physics, University of Maryland, College Park, MD 20742}
\date{\today}
\abstract{We study possible extensions of the Twin Higgs model that solve the Hierarchy problem and simultaneously address problems of the large- and small-scale structures of the Universe. Besides naturally providing dark matter (DM) candidates as the lightest charged twin fermions, the twin sector contains a light photon and neutrinos, which can modify structure formation relative to the prediction from the $\Lambda$CDM paradigm. We focus on two viable scenarios. First, we study a Fraternal Twin Higgs model in which the spin-3/2 baryon $\hat{\Omega}\sim(\hat{b}\hat{b}\hat{b})$ and the lepton twin tau $\hat{\tau}$ contribute to the dominant and subcomponent dark matter densities. A non-decoupled scattering between the twin tau and twin neutrino arising from a gauged twin lepton number symmetry provides a drag force that damps the density inhomogeneity of a dark matter subcomponent. Next, we consider the possibility of introducing a 
twin hydrogen atom $\hat{H}$ as the dominant DM component. After recombination, a small fraction of the twin protons and leptons remains ionized during structure formation, and their scattering to twin neutrinos through a gauged U$(1)_{B-L}$ force provides the mechanism 
that damps the density inhomogeneity. Both scenarios realize the Partially Acoustic dark matter (PAcDM) scenario and explain the $\sigma_8$ discrepancy between the CMB and weak lensing results. Moreover, the self-scattering neutrino behaves as a dark fluid that enhances the size of the Hubble rate $H_0$ to accommodate the local measurement result while satisfying the CMB constraint. For the small-scale structure, the scattering of $\hat{\Omega}$'s and $\hat{H}$'s through the twin photon exchange generates a self-interacting dark matter (SIDM) model that solves the mass deficit problem from dwarf galaxy to galaxy cluster scales. Furthermore, when varying general choices of the twin photon coupling, bounds from the dwarf galaxy and the cluster merger observations can set an upper limit on the twin electric coupling.}
\maketitle
\section{Introduction}
We study a non-minimal dark sector motivated by both Naturalness and cosmology considerations and explore its potential. By doing so, we provide a solution to the little hierarchy problem and, simultaneously, to various cosmological structure anomalies suggested by the current data related to the large- and small-scale structure of the universe. The existence of these issues may have revealed an intriguing clue to the nature of dark matter. 

The Twin Higgs mechanism \cite{Chacko:2005pe,Chacko:2005un,Batra:2008jy} provides a solution to the little hierarchy problem in a \emph{hidden naturalness} manner. The solution evades strong constraints from the Large Hadron Collider (LHC) on top-partners that are charged under Standard Model (SM) color by furnishing a hidden SM-like sector, in which the SM-neutral twin top is involved in stabilizing the Higgs mass. There have been several studies on formulating an ultraviolet completion of the model \cite{Chang:2006ra,Falkowski:2006qq,Craig:2013fga,Geller:2014kta,Craig:2014roa,Craig:2014aea,Barbieri:2015lqa,Low:2015nqa,Csaki:2015gfd,Craig:2016kue,Yu:2016bku} and on the collider phenomenology related to the twin particle spectrum
\cite{Burdman:2014zta, Craig:2015pha, Curtin:2015fna,Cheng:2015buv, Chacko:2015fbc}. 
The existence of the mirror sector also provides a non-minimal dark sector containing stable charged fermions and twin gauge bosons, which introduces various applications to cosmology. Previous works on twin cosmology mainly focused on the thermal history of the dark matter candidates \cite{Craig:2015xla,Farina:2015uea,Garcia:2015toa,Farina:2016ndq} and signatures in the (in-)direct detection experiments \cite{Garcia:2015loa,Freytsis:2016dgf}. In this work, we explore the physics of structure formation in the context of the Twin Higgs model.

A dark sector that contains a SM-like particle spectrum has the potential to extend the cold collisionless dark matter paradigm in a way that resolves important cosmological issues \cite{Foot:2014mia,Foot:2014uba,Foot:2016wvj}. In the twin sector, the dark matter candidates are the lightest charged baryon and lepton, which scatter with each other via twin photon exchange during the halo formation. 
The twin sector also contains a light twin neutrino, whose existence affects the expansion rate of the universe and hence shifts $H_0$, the value of the Hubble expansion rate today. If the twin sector is extended to include an efficient scattering between the twin neutrino and charged fermions during the structure formation time, a dark acoustic oscillation exists, which damps the dark matter power spectrum and alters the large-scale structure.

Interestingly, these adjustments to the dark matter structure formation in fact furnish solutions to the existing inconsistencies between the $\Lambda$CDM prediction and both the large- and small-scale structure observations. For many years, the well-accepted $\Lambda$CDM paradigm has provided an excellent fit to cosmological data on large scales, although there had been several long-standing problems on small scales, including the core-vs-cusp \cite{massdeficit, Flores:1994gz} and too-big-to-fail problems \cite{2011MNRAS.415L..40B}. With the advent of higher-precision measurements on large scales, however, the large-scale results have entered into tension with $\Lambda$CDM as well. In particular, there is a $\sim 3\sigma$ discrepancy between the value of today's Hubble rate $H_0$ obtained from a fit to the CMB and baryon acoustic oscillation (BAO) data \cite{Ade:2015xua} and the higher results from local measurements \cite{Riess:2011yx, 2013ApJ...766...70S, Riess:2016jrr,DiValentino:2016hlg, Bernal:2016gxb}. 
Further, the inferred value of $\sigma_8$ (roughly speaking the amplitude of matter density fluctuations at a scale of $8h^{-1}$ Mpc) is in $2-3\sigma$ tension \cite{Heymans:2013fya, Ade:2013lmv, MacCrann:2014wfa, Poulin:2016nat} with the lower values from direct measurements by the weak lensing survey \cite{Fu:2014loa}. Resolving these anomalies would require a paradigm that generically reduces the value of $\sigma_8$ and enhances $H_0$ as compared to the $\Lambda$CDM model in a consistent way.

One attempt to raise $H_0$ from the $\Lambda$CDM prediction is to introduce additional dark radiation (DR) to increase the energy density. 
Once the stringent CMB constraints are taken into account, however, such a solution comes at the cost of increasing the matter power spectrum, which exacerbates the $\sigma_8$ problem \footnote{For example, see the $\sigma_8-H_0$ contours in Fig.~33 of \cite{Ade:2015xua}.}. A plausible solution is to have the dark radiation, which enhances $H_0$, also act to damp the dark matter power spectrum so that the size of $\sigma_8$ gets reduced to agree with the weak lensing result \cite{Buen-Abad:2015ova}. One can consider coupling all the dark matter particles to the dark radiation. For this scenario, the full DM-DR system undergoes dark acoustic oscillations, and hence all dark matter components are subjected to the same damping. Such a proposal would require a well-chosen small DM-DR coupling, which results in a DM-DR scattering that is slightly inefficient when compared to the Hubble expansion. Consequently, a numerical study is necessary to obtain the correct $\sigma_8$ suppression \cite{Lesgourgues:2015wza,Ko:2016uft,Ko:2016fcd}. It is because of this slightly inefficient scattering process that we refer to this setup as the Quasi-Acoustic Dark Matter (QuAcDM) scenario. In the Twin Higgs model, such a scenario can be realized by gauging the twin B$-$L symmetry. Here the twin neutrino plays the role of the additional dark radiation, and its scattering to the dark matter (twin baryon and charged lepton) damps the dark matter power spectrum, solving the $\sigma_8$ problem.

Alternatively, one can consider a scenario where only a subcomponent of the total dark matter couples to the dark radiation. In a well-motivated general mechanism that was recently introduced in \cite{Chacko:2016kgg}, one can allow the DM-DR scattering to be highly efficient. The Partially Acoustic Dark Matter (PAcDM) is a robust framework that effectively resolves both the $\sigma_8$ and $H_0$ large-scale structure anomalies in a natural way. It assumes the presence of tightly coupled dark radiation and supposes that the dark matter mass density is composed of two components, a cold and collisionless dominant one ($\chi_1$) and a cold subdominant one ($\chi_2$) that is tightly coupled to the dark radiation. The success of this framework hinges on the feature that both the self-interaction of the dark radiation and the DR-$\chi_2$ interaction remain efficient throughout the radiation domination phase and for a significant portion of the structure formation era. For this reason, one can perform an analytical estimation of the $\sigma_8$ suppression in the tightly coupled limit, as we will discuss in Sec.~\ref{sec:LSS}.

In this framework, the Hubble parameter anomaly can be reconciled by suitably fixing the amount of tightly coupled dark radiation. Further, for the 
$\sigma_8$ anomaly, if the relevant modes enter the horizon before matter-radiation equality, the interaction between the dark radiation and $\chi_2$ restricts the growth of density perturbations, subsequently decreasing the growth of fluctuations in the collisionless DM $\chi_1$. This is the case for modes sensitive to the $\sigma_8$ measurement, and hence the discrepancy can be resolved by an appropriate choice of the amount of subdominant DM, reducing the $\sigma_8$ value to match the observed deviation. 
Furthermore, the reduced growth of the matter power spectrum in this scenario results in a minor correction to the gravity perturbation during the CMB time, yielding a smaller change of the CMB spectrum as compared to the QuAcDM case. Hence, future precision CMB studies may be able to distinguish these two classes of models.

\begin{figure} 
\includegraphics[width=6cm]{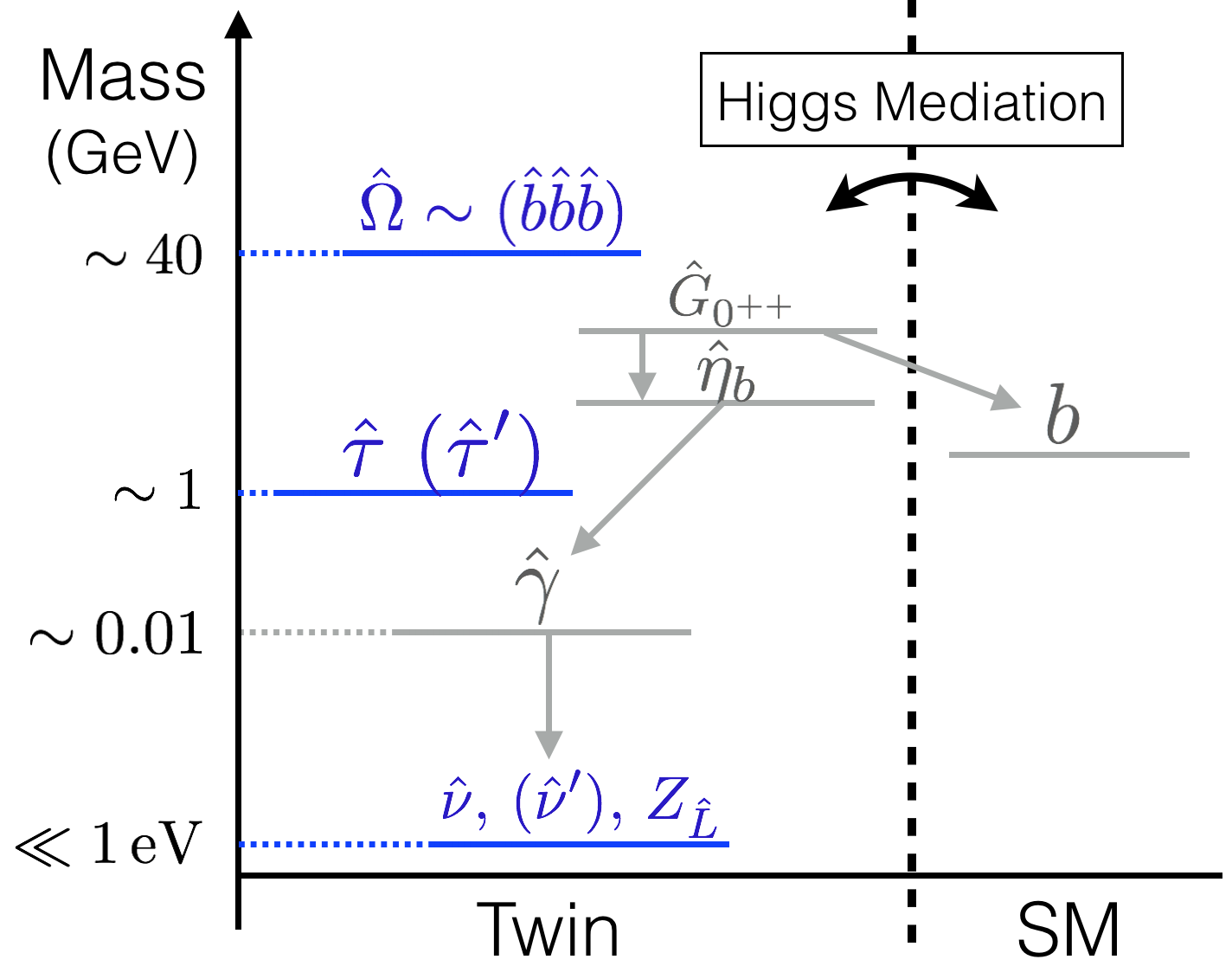}\quad\includegraphics[width=9.5cm]{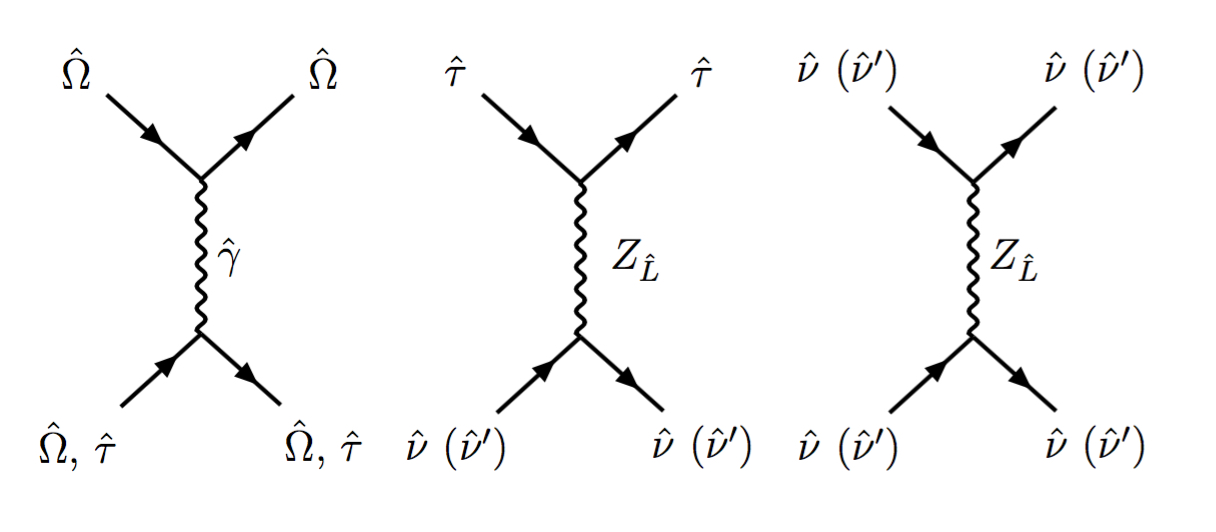}
\caption{Left: A representation of the particle spectrum in the extended Twin Higgs model under consideration. The blue (gray)-colored particles correspond to the stable (unstable) members of the spectrum. The gray arrows indicate the decay products of the unstable particles. Further, the primed fields correspond to the set of specific anomaly compensators used in this paper. 
Right: The set of dominant processes involved in the solution of the large- and small-scale structure anomalies. The first Feynman diagram represents the dominant process relevant for the self-interacting dark matter scenario through the exchange of twin photons. The second diagram corresponds to the relevant scattering for the partially acoustic oscillation scenario, and the third one is the process that keeps the dark radiation a tightly coupled fluid. 
}\label{fig:scattering}
\end{figure}

We focus on the PAcDM scenario here in the context of the Twin Higgs model. Our particular realization is obtained by gauging either the twin lepton number symmetry U$(1)_L$ or the twin U$(1)_{B-L}$. In the U$(1)_L$ case, the heavy twin lepton scatters with the twin neutrino and plays the role of $\chi_2$, while the twin baryon behaves as cold collisionless dark matter $\chi_1$ throughout structure formation (Fig.~\ref{fig:scattering}). In the U$(1)_{B-L}$ case, the twin hydrogen behaves as $\chi_1$, while the ionized twin proton plays the role of $\chi_2$. The scattering between light twin particles also renders the light degrees of freedom a tightly coupled fluid, which gives an extra contribution to $\Delta N_{\text{eff}}$, and suitably solves the $H_0$ problem, while satisfying a weaker CMB constraint \cite{Friedland:2007vv, Baumann:2015rya, fluidbound}.

In addition to the large-scale structure anomalies, there are several long-standing puzzles on small scales related to the structure of dark matter halos that cannot be addressed by the collisionless dark matter models. In particular, direct observations of dwarf galaxies ($\sim$ kpc size) and galaxy clusters 
($\sim$ Mpc size) indicate lower dark matter masses in the inner regions of these objects than those predicted by N-body simulations with non-interacting DM. 
Although this anomaly may potentially be explained by lack of baryon interaction in the simulations \cite{2012MNRAS.422.3081M,2012MNRAS.422.1231G}, none of the proposed solutions so far are able to cover such a broad range of halo sizes simultaneously \footnote{See \cite{Vogelsberger:2015gpr} for a review of current status on the small scale structure problems.}. One attractive solution to the mass deficit problem on all halo scales is to suppose that the dark matter is self-interacting through a light mediator. As we show in this work, the charged twin baryon provides a plausible realization of this self-interacting dark matter (SIDM) scenario \cite{Spergel:1999mh,Kaplinghat:2015aga} through an $\mathcal{O}(10)$ MeV-scale twin photon. Alternatively, the dark matter self-interaction can be realized for 
a DM particle with an extended geometrical size for the scattering, e.g. for atomic DM. As we will show, the formation of twin hydrogen gives a natural realization of atomic DM \cite{Goldberg:1986nk,Kaplan:2009de,Kaplan:2011yj,CyrRacine:2012fz,Cline:2013pca} and provides solutions to the mass deficit problem from dwarf galaxy to galaxy cluster scales \cite{Boddy:2016bbu}.

The rest of this paper is organized as follows. In Sec.~\ref{sec:model}, we describe the Twin Higgs model that contains the necessary ingredients to solve both the large- and small-scale structure puzzles at the same time. The model is based on the Fraternal Twin Higgs model \cite{Craig:2015pha} but has a gauged twin lepton number symmetry that generates the $\sigma_8$ damping. 
In Sec.~\ref{sec:LSS}, we explain how this model serves as a realization of the PAcDM framework to resolve the $(H_0,\sigma_8)$ anomalies. We give an analytical description of the partially acoustic oscillation and calculate the required mass ratio between the stable twin baryon and lepton that solves these problems. In Sec.~\ref{sec:SIDM} we discuss the solution from the Twin Higgs model to the mass deficit problem and calculate the mass of the twin photon necessary to resolve it. At the end, when we admit more general choices of the twin electric coupling, we demonstrate how the dwarf galaxy and cluster merger observations enable us to set an upper bound on this coupling. In Sec.~\ref{sec:twinH}, we discuss the more attractive solution to the cosmological structure formation problems with the twin hydrogen playing the role of the dominant DM component. An eminent virtue of this model is that it requires neither the breaking of twin electromagnetism nor the presence of anomaly compensators, which suffer from strong experimental constraints. Moreover, it successfully accommodates two generations of fermions and does not compel one to introduce additional mass scales, thus keeping the twin gauge bosons massless. We conclude in Sec.~\ref{sec:conclusion}.

\section{The Extended Fraternal Twin Higgs Model}\label{sec:model}

\subsection{Asymmetric Dark Matter and Dark Fluid}
We investigate a dark sector motivated by the Twin Higgs model, which contains a twin top Yukawa and mirror gauge symmetries 
SU$(3)'_c\times$SU$(2)'_L\times$U$(1)_{Y'}$ with SM-like couplings. In this framework, the SM Higgs arises as a pseudo-goldstone boson from a global SU$(4)$ symmetry breaking, which then enjoys protection from various radiative corrections due to the approximate $\mathbb{Z}_2$ symmetry. Since the Higgs mass receives subdominant corrections from the Yukawa interactions of light twin fermions, all Yukawa couplings except for the twin top coupling are mildly constrained by Naturalness. For this reason, one can simplify the twin fermion spectrum by including only the third-generation fermions. In this Fraternal Twin Higgs model \cite{Craig:2015pha}, the approximately $\mathbb{Z}_2$-symmetric gauge and top Yukawa couplings can adequately stabilize the Higgs mass, while the smaller number of light fermions is able to more flexibly satisfy the restrictive $\Delta N_{\text{eff}}$ bound from the CMB. 

Since we mainly focus on the thermal history of the twin sector below a temperature of $\mathcal{O}(10)$ GeV, the twin particles relevant for this discussion are the spin-$3/2$ twin baryon $\hat{\Omega}=(\hat{b}\hat{b}\hat{b})$, the twin tau $\hat{\tau}$, the twin neutrino $\hat{\nu}$ (with both chiralities), and the twin photon $\hat{\gamma}$. Their mass spectrum is similar to the SM spectrum due to the approximate $\mathbb{Z}_2$ symmetry. If we take the ratio between the twin and SM electroweak symmetry breaking scales, $f$ and $v$, to be $f/v=3$, which corresponds to a minor $2(v/f)^2\simeq 20\%$ tuning and satisfies current constraints on the Higgs coupling \cite{Craig:2015pha}, the twin top and twin gauge bosons feature masses larger than the SM values by a factor of three.
 
The approximately $\mathbb{Z}_2$-symmetric Yukawa couplings of light twin fermions $\hat{b}$ and $\hat{\tau}$, which result in small corrections to the Higgs mass, can be modified from the SM values, leading to some arbitrariness in the determination of the twin particle masses. We will therefore set the masses according to the solutions of the large- and small-scale structure problems. The relevant parameters are the twin photon mass $m_{\hat{\gamma}}$, the twin baryon mass $m_{\hat{\Omega}}$, and the ratio of the twin tau mass density to the total dark matter density. We write the latter two quantities as
\begin{equation}\label{eq:masses}
m_{\hat{\Omega}}\simeq 3m_{\hat{b}}+5\hat{\Lambda},\quad r\equiv\frac{m_{\hat{\tau}}}{m_{\hat{\Omega}}+m_{\hat{\tau}}}.
\end{equation}
Here, the contribution from the SU$(3)'_c$ confinement scale $\hat{\Lambda}$ is due to an approximation that comes from the lattice result in Ref.~\cite{Farchioni:2007dw} for a spin-$3/2$ baryon in the single-flavor case \cite{Garcia:2015toa}. When solving the mass deficit problem through the $\hat{\Omega}$ self-scattering, the discussion in Sec.~\ref{sec:SIDM} will demand mass ranges $10\lsim m_{\hat{\gamma}}\lsim 20$ MeV and $10\lsim m_{\hat{\Omega}}\lsim40$ GeV. On the other hand, for the discussion in Sec.~\ref{sec:LSS}, 
the appropriate damping of the $\sigma_8$ result will call for a mass ratio of $r\simeq 2.5\%$ and the existence of relativistic twin neutrinos. 
To simplify the discussion of thermal history, we assume the twin neutrinos to be massless during structure formation and focus on the following mass parameters:
\begin{equation}\label{eq:benchmark}
m_{\hat{\Omega}}=40\,\text{GeV},\quad m_{\hat{\gamma}}=10\,\text{MeV},\quad r = 2.5\%.
\end{equation}
These imply that $m_{\hat{\tau}}\simeq 1$ GeV. The $\hat{\Omega}$ mass further implies that $m_{\hat{b}}\simeq 5$ GeV for the twin confinement scale of $\hat{\Lambda}\simeq 5$ GeV, which comes from the two-loop RG running of $\mathbb{Z}_2$-symmetric QCD couplings at the cutoff scale that we assume to be $5$ TeV \cite{Craig:2015pha}. The $\mathbb{Z}_2$-breaking Yukawa couplings of $\hat{b}$ and $\hat{\tau}$ yield a cutoff ($\Lambda$)- dependent correction to the Higgs mass,
$\delta m_h^2\simeq\frac{\Lambda^2}{4\pi^2}(\Delta y_{b}^2+\Delta y_{\tau}^2)$, where $\Delta y_b^2\equiv 3(y_{\hat{b}}^2-y_{b}^2)$ and $\Delta y_{\tau}^2\equiv (y_{\hat{\tau}}^2-y_{\tau}^2)$ \cite{Craig:2015pha}. For $\Lambda=5$ TeV, we find no significant tuning of the Higgs mass, $\delta m_h^2\simeq (0.27\,m_h)^2$.

Assuming an unbroken twin electric symmetry, the twin electrically charged $\hat{\Omega}$ and $\hat{\tau}$ particles are stable and can serve as dark matter candidates in this setup. The relic abundance of $\hat{\Omega}$ can be generated through a similar baryogenesis mechanism as in the SM sector. With a small difference in either the CP violation or first order phase transition, we can achieve a different baryon asymmetry $Y_{\Delta \hat{B}}\simeq Y_{\Delta B}/8$ relative to the SM. This generates the observed dark matter density. If we suppose that the twin-sector remains charge neutral from the twin baryogenesis, then given the absence of other stable charged particles, we expect the number of $\hat{\Omega}$ in the late-time Universe to coincide with the number of $\hat{\tau}$. Besides the dark matter particles, light hadrons like the $0^{++}$ glueball, of mass $m_{\hat{G}_{0^{++}}}\simeq 6.9\hat{\Lambda} = 35$ GeV, and the pseudo-scalar bottomonium, of mass $m_{\hat{B}_{0^{-+}}}\simeq 2(m_{\hat{b}}+\hat{\Lambda})=20$ GeV, decay quickly into the SM $b\bar{b}$ or twin photons when they become non-relativistic. Hence, we do not consider them in the discussion of structure formation.

Let us now discuss some observational constraints relevant to the dark sector. We first turn to the direct detection constraint on the dominant dark matter component $\hat{\Omega}$. This is determined by the spin-independent cross section of $\hat{\Omega}\,p\to\hat{\Omega}\,p$ through the Higgs portal exchange, given by \cite{Craig:2015xla}
\begin{equation}
\sigma_h\simeq\frac{1}{\pi}\left(\frac{3\,y_{\hat{b}}v}{\sqrt{2}f}\right)^2g_{hp}^2\frac{\mu_{N\hat{\Omega}}^2}{m_h^4},
\end{equation} 
where $\mu_{N\hat{\Omega}}$ is the reduced mass of the $\hat{\Omega}$-nucleon system and $g_{hp}= 1.2\times 10^{-3}$ \cite{Craig:2015xla,Crivellin:2013ipa} gives the effective Higgs coupling to nucleons. Since the momentum transfer in the scattering is much smaller than the inverse of the $\hat{\Omega}$ radius, we assume that the Higgs mediation is dominated by the coherent scattering to three $\hat{b}$'s in the bound state,
which includes a factor of $3^2$ in the cross section. Taking $m_{\hat{b}}=5$ GeV, this expression gives $\sigma_h\simeq3.4\times 10^{-47}$ cm$^{2}$. This value falls below the current bound $\simeq 1.0\times 10^{-46}$ cm$^2$ ($90\%$ CL) from the LUX experiment \cite{Akerib:2016vxi} at $40$ GeV dark matter mass, but the cross section lies within the sensitivity of the proposed LZ experiment \cite{Akerib:2015cja}. As is discussed in \cite{Garcia:2015toa}, the $\hat{\Omega}$-Higgs coupling can also be generated from a scalar glueball exchange. The resulting cross section may be comparable to the Higgs mediation result, but a concrete result relies on a future lattice study.

\subsection{Dark Matter Self-Interaction and Dark Matter-Dark Fluid Scattering}
Let us next turn to the dark matter interactions relevant to the formation of large- and small-scale structure. For the small-scale structure case, the model in question assumes that the dark matter particles $(\hat{\Omega},\,\hat{\tau})$ carry twin electric charges and are endowed with self-couplings; hence, they elastically self-scatter. Although this self-scattering does not affect the linear evolution of large-scale structure \cite{Cyr-Racine:2015ihg}, it can influence the dark matter structure formation.
As we show later, we choose the same value for the twin and SM fine structure constants, $\hat{\alpha}=\alpha$, and a twin photon mass $m_{\hat{\gamma}}\sim 10$ MeV that softly breaks the $\mathbb{Z}_2$ symmetry. The photon mass enables us to generate the appropriate velocity-dependent cross section that explains small-scale structure anomalies from dwarf to galaxy cluster scales. If the U$(1)_{\hat{Y}}$-breaking spurion $m_{\hat{\gamma}}$ carries a fractional charge, then $\hat{\Omega}^{\pm}$ and $\hat{\tau}^{\mp}$ can be easily made to be long-lived when compared to the cosmological time scale. Since $m_{\hat{\gamma}}$ is larger than the binding energy $m_{\hat{\tau}}\hat{\alpha}^2\simeq40$ keV, the two particles do not form a bound state through the $\hat{\gamma}$ exchange. We also note that $\hat{\Omega}$ can also self-scatter via the exchange of twin mesons. However, this corresponds to a mediation scale that is above a GeV, in which case the resulting self-interaction is too weak to explain the anomalies.

Having a new mass scale $m_{\hat{\gamma}}$ in the twin sector complicates the UV-completion of the model. Since the hyper charge U$(1)_Y$ gives a negligible contribution to Higgs tuning, if the goal of the model is to only solve one of the small-scale structure problems, one can solve the dwarf anomaly by a $\mathbb{Z}_2$-breaking coupling $\hat{\alpha}\sim 10^{-2}\alpha$ and assume the cluster anomaly to be resolved by some SM baryonic effect. However, in order to demonstrate the potential of the twin sector in addressing the structure formation issues, we will still aim for a solution to the small-scale structure anomalies on all scales and focus on the massive twin photon scenario. We will also discuss an alternative SIDM scenario in Sec.~\ref{sec:twinH} that does not require a massive $\hat{\gamma}$. In this scenario, the twin hydrogen plays the role of the SIDM, and the additional velocity dependence in the scattering cross section 
in both the elastic and inelastic scattering processes resolves both the dwarf and cluster anomalies, once the hyperfine structure of the twin atoms is taken into account.

Turning to the large-scale structure in the $\hat{\Omega}$-$\hat{\tau}$ scenario, we find that in order to address the $\sigma_8$ puzzle in the PAcDM framework, we introduce a non-decoupled interaction between the subdominant dark matter $\hat{\tau}$ and dark radiation $\hat{\nu}$ that acts to damp the matter density contrast. Any such interaction between $\hat{\nu}$ and $\hat{\tau}$ but not $\hat{\nu}$ and $\hat{\Omega}$ can serve this purpose. To provide a specific scenario, we implement this interaction by gauging the twin lepton number symmetry and assuming that U$(1)_{\hat{L}}$ is preserved throughout structure formation. There is then an efficient scattering $\hat{\tau}\hat{\nu}\to\hat{\tau}\hat{\nu}$ through a massless $\hat{Z}_{L}$ mediator.

Gauging the U$(1)_{\hat{L}}$ symmetry results in local gauge anomalies. 
We can keep the U$(1)_{\hat{L}}$ symmetry anomaly-free during structure formation by introducing anomaly compensators. For example, one way to achieve this is to include twin leptons $\hat{l}'^T_R=(\hat{\nu}',\,\hat{\tau}')_R\sim (1,2,0,1)$, $\hat{\tau}'_L\sim (1,1,-1/2,1)$ and $\hat{\nu}'_L\sim(1,1,1/2,1)$ charged under the twin SU$(3)'_c\times$SU$(2)'_L\times$U$(1)_{\hat{Y}}\times$U$(1)_{\hat{L}}$\footnote{Instead of having anomaly compensators simply as chirality-flipped twin fermions \cite{FileviezPerez:2010gw}, here we assign different U$(1)_{\hat{Y}}$ charges to the neutrino compensators, so that they do not introduce vectorized neutrino masses, and the twin photon can decay into $\hat{\nu}'$'s before BBN. Having the twin photon decay into dark radiation can avoid the stringent direct detection constraints as compared to the decay into SM particles through a kinetic mixing.}. 
We assume that the tau compensator, $\hat{\tau}'$, obtains a Yukawa mass from the term 
$y_{\hat{\tau}}\bar{\hat{l}}'_R\hat{H}\hat{\tau}'_L$ that is slightly heavier than that of $\hat{\tau}$, 
so that $\hat{\tau}'$ decays quickly into $\hat{\tau}\bar{\hat{\nu}}\hat{\nu}'$ when it becomes nonrelativistic. 
We also assume that the twin neutrino compensator $\hat{\nu}'$ remains massless just like $\hat{\nu}$.  
This particle provides an additional contribution to $\Delta N_{\text{eff}}^{\text{scatt}}$ and consequently helps to explain a higher value of $H_0$ from the local measurements. 
 A potential cause for concern is the allowed decay of $\hat{\tau}$ into the neutral ($\hat{\nu}$) and charged ($\hat{\nu}'$) neutrinos, which comes from a dimension-10 operator $(\bar{\hat{l}}_L\hat{H}\hat{\nu}_R)(\bar{\hat{\nu}}'_L\hat{\nu}_R)(\bar{\hat{\nu}}'_L\hat{\nu}_R)$. However, this concern is eliminated if the mediation scale is above a TeV. In our discussion of the acoustic oscillation, we take the size of the U$(1)_{\hat{L}}$ coupling to be $g_{\hat{L}}\gsim 10^{-4}$ to ensure that the $\hat{\tau}$-neutrino scattering rate is always larger than the Hubble expansion rate.

Further, it turns out that in order to evade the stringent bound from searches for a fifth force in the SM sector (see \cite{Harnik:2012ni} for a review of the constraints), we are led to retain the SM U$(1)_L$ as un-gauged, otherwise the same anomaly compensators in the SM suffer from stringent collider constraints. Since the U$(1)_{\hat{L}}$ interaction only affects the twin Higgs mass at two-loop level, this minor $\mathbb{Z}_2$ breaking has a negligible effect on the naturalness of the electroweak scale. In the discussion of twin hydrogen DM in Sec.~\ref{sec:twinH}, the scattering between the ionized twin atom and $\hat{\nu}$ is given by a gauged U$(1)_{B-L}$ symmetry. 
Since U$(1)_{B-L}$ is anomaly-free, there is no need to introduce the unattractive anomaly compensators. Hence, it is easier to UV-complete the model by gauging U$(1)_{B-L}$ in both sectors, and break the SM U$(1)_{B-L}$ through the same $\mathbb{Z}_2$ breaking as in the Higgs potential.

Another possible cause for concern is the Weibel plasma instability. Inside the halo, the twin tau in the dark fluid behaves like a charged plasma, and there may be potential constraints on the U$(1)_{\hat{L}}$ coupling from the plasma instability \cite{Heikinheimo:2015kra,Agrawal:2016quu}. However, since the twin tau density is only $2.5\%$ of the overall dark matter density, we do not expect this bound to be strong. Incidentally, we should note that the precise bound has not yet been formulated and is currently still under construction.

In our PAcDM scenario, the $\hat{\tau}-\hat{\nu}$ scattering is highly efficient, rendering the dark radiation $(\hat{\nu},\hat{\nu}', Z_{\hat{L}}$) a tightly coupled fluid. Like free-streaming radiation, this dark fluid contributes to the effective number of neutrino species $\Delta N_{eff}^{scatt}$. However, being self-interacting, the fluid is subject to a weaker CMB constraint when compared to the free streaming case with $\Delta N_{eff}^{scatt}<1.06$ $(2\sigma)$ \cite{Baumann:2015rya, fluidbound}. This feature has the effect of freeing up space to accommodate $\Delta N_{eff}^{scatt}\simeq 0.4-1$, which furnishes a solution to the $H_0$ problem \cite{Riess:2016jrr}. To determine the $\Delta N_{eff}^{scatt}$ in our model, we refer to the state of the Universe around the kinetic decoupling time between the SM and twin sectors. 
Kinetic equilibrium between the two sectors is maintained by the Higgs mediation, which decouples around the GeV scale. 
Immediately after the decoupling, the twin sector contains the relativistic particles $(\hat{\gamma},\,Z_{\hat{L}},\,\hat{\nu},\,\hat{\nu}')$. As soon as the temperature drops to $T \lsim 10$ MeV, the twin photon $\hat{\gamma}$ decays into $\hat{\nu}'$'s, avoiding the stringent direct detection constraints it would suffer if it were to instead decay into SM particles through a kinetic mixing before Big-bang Nucleosynthesis \cite{Kaplinghat:2013yxa}.
Now, after the twin photons decay, the twin sector is left with $(\hat{\nu},\,\hat{\nu}',\,Z_{\hat{L}})$, which contribute an overall $\Delta N_{eff}^{scatt}\simeq 0.46$. This size of $\Delta N_{eff}^{scatt}$ is large enough to adequately enhance the Hubble rate, solving the $H_0$ problem.

An alternative solution to the $(H_0, \sigma_8)$ problems is to gauge the anomaly-free   
U$(1)_{B-L}$ symmetry instead of gauging the twin lepton number. This is a realization of the QuAcDM framework. In contrast to the PAcDM for which only the subdominant dark matter component undergoes dark acoustic oscillations, here the acoustic oscillations are experienced by the full DM-DR system, namely by both $\hat{\Omega}$ and $\hat{\tau}$ interacting with the dark radiation. 
These damp the power spectrum with a weak U$(1)_{B-L}$ coupling. If we invoke the result in Ref.~\cite{Buen-Abad:2015ova, Lesgourgues:2015wza}, we can reduce the size of $\sigma_8$ to the desired value by choosing $\hat{\alpha}_{B-L}\sim 10^{-9.8}$ for a $10$ GeV-scale dark matter mass. 

\section{Large-scale Structure: Twin Lepton with Acoustic Oscillations}\label{sec:LSS}
\begin{figure}
\begin{center}
\includegraphics[width=10cm]{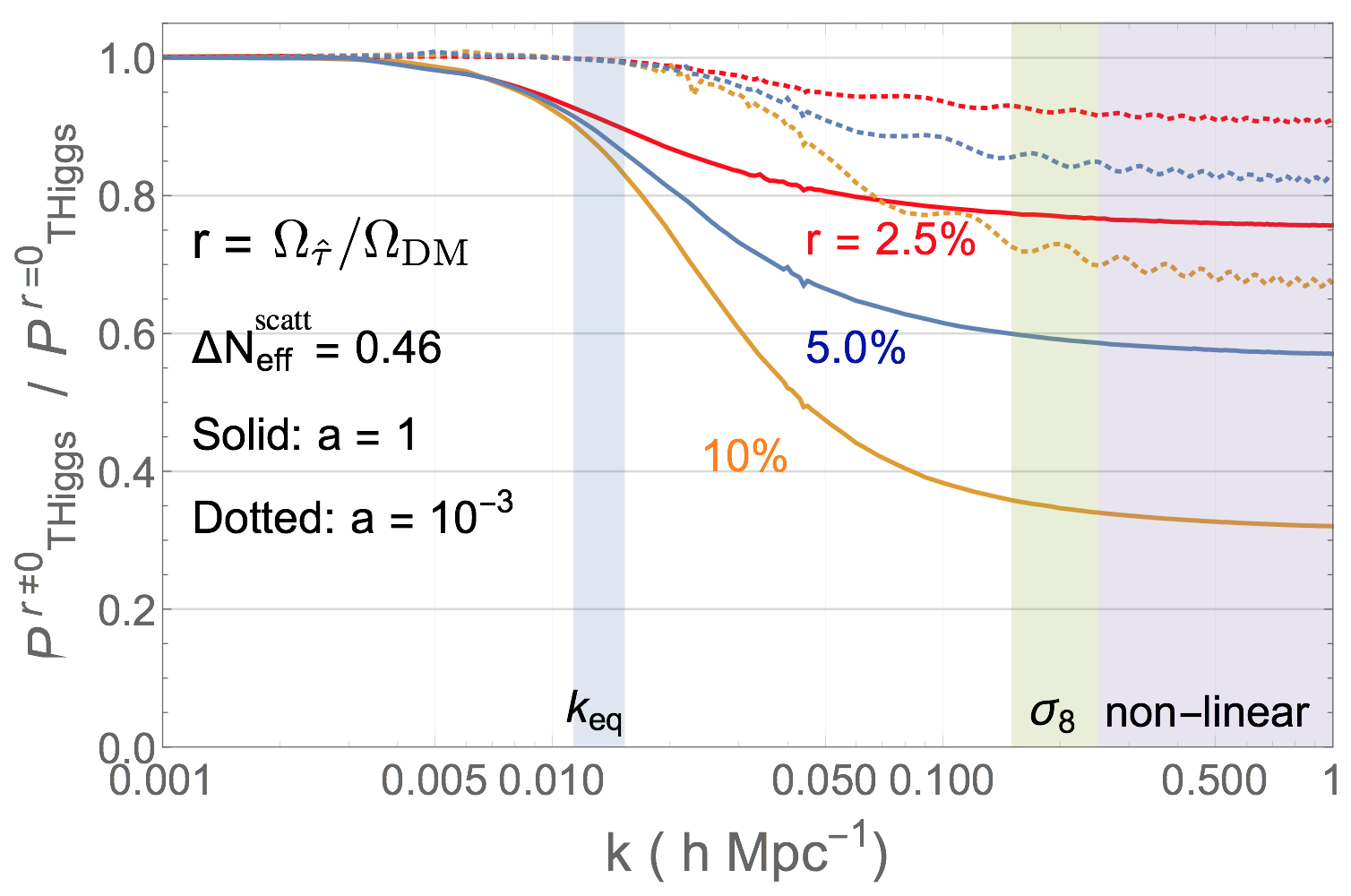}
\end{center}
\caption{Ratio of the dark matter power spectrum between the $r\neq 0$ and $r=0$ cases, both with 
$N_{\text{eff}}^\text{scatt} = 0.46$. In the Twin Higgs setup discussed in this work, the dark matter ratio 
$r$ is given by $r=m_{\hat{\tau}}/(m_{\hat{\Omega}}+m_{\hat{\tau}})$. In the plot, the solid (dashed) curves are obtained by numerically solving the linear evolution equations described in Sec.~\ref{sec:LSS}, all in the tight coupling limit and assuming no anisotropic stress. Results for different values of $r$ are labelled in different colors, while earlier ($a = 10^{-3}$) and late ($a=1$) times are indicated by dotted and solid lines, respectively. Also see Ref.~\cite{Chacko:2016kgg} for more details.}
\label{fig:psratio}
\end{figure}
\begin{figure}
\begin{center}
\includegraphics[width=10cm]{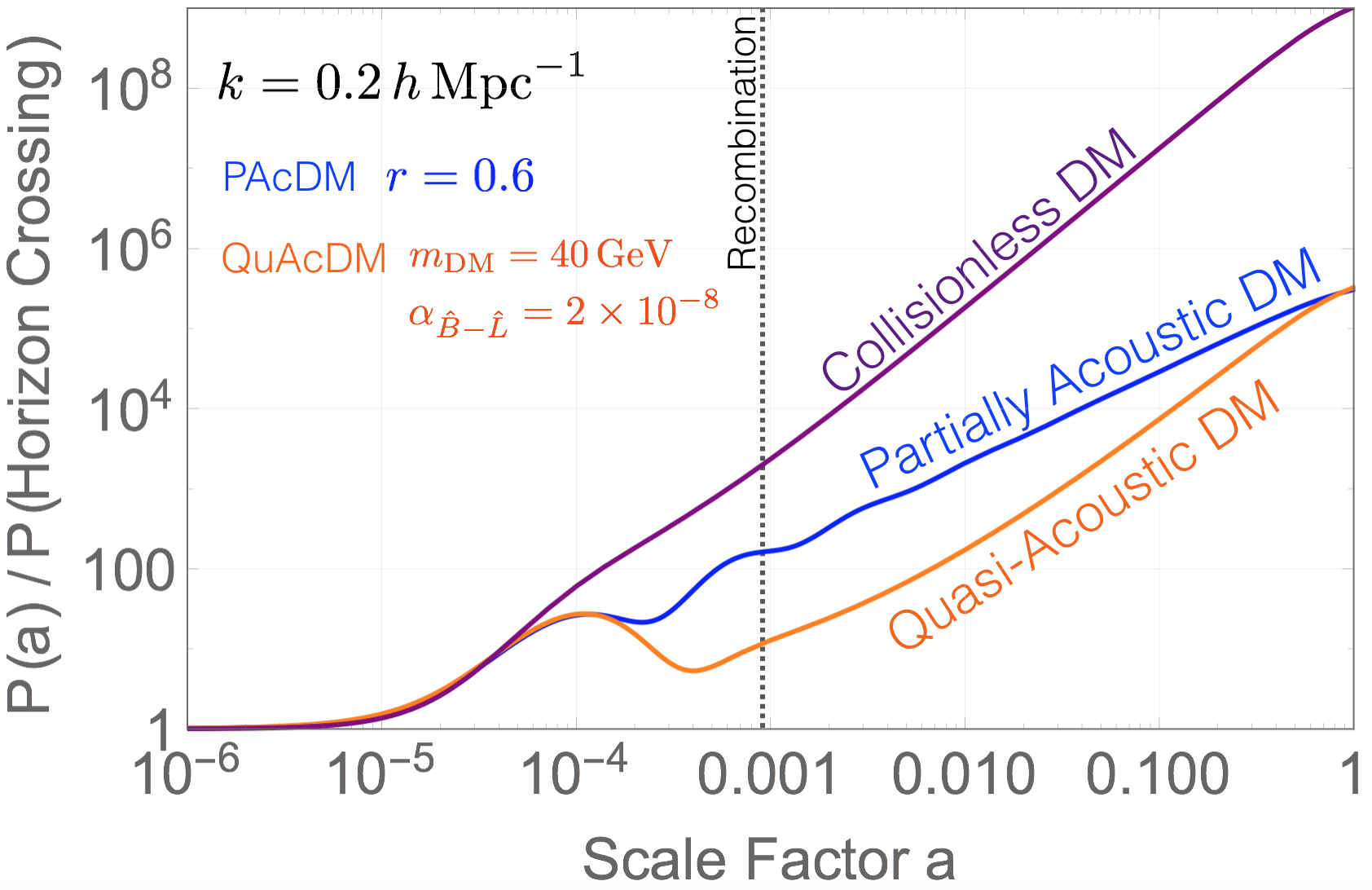}
\end{center}
\caption{An example of the evolution of dark matter power spectrum in the Cold dark matter (purple), Partially Acoustic dark matter (PAcDM, blue), and Fully Acoustic dark matter (QuAcDM, orange) cases. In the QuAcDM
scenario, which corresponds to the case when both $\hat{\Omega}$ and $\hat{\tau}$ interact with the dark radiation, the oscillation delays the linear growth of the density contrast during the matter domination phase. This feature results in a suppression of the matter power spectrum. In the PAcDM scenario, which corresponds to the case when only $\hat{\tau}$ scatters with the dark radiation, the slower growth of the power spectrum allows a smaller deviation from the CDM case during the CMB time ($a\sim 10^{-3}$) and the same suppression of power spectrum today ($a\simeq 1$) as in the QuAcDM case. In order to illustrate the idea, we choose parameters that give a large $\sigma_8$ suppression.}
\label{fig:cartoon}
\end{figure}

The presence of a cold dark matter component $\hat{\Omega}$ and a subcomponent dark matter $\hat{\tau}$ that couples to the self-scattering radiation $(\hat{\nu},\,\hat{\nu}',\,Z_{\hat{L}})$, modifies the values of $(H_0,\,\sigma_8)$ 
relative to the $\Lambda$CDM model. 
The self-scattering radiation contributes an overall $\Delta N_{eff}^{scatt}\simeq 0.46$ that serves to reconcile the values of $H_0$ between the local and CMB measurements \cite{Riess:2011yx}. Moreover, during the matter-dominated era, the dark fluid-$\hat{\tau}$ scattering generates a dark acoustic oscillation that delays the structure formation of $\hat{\tau}$.  As we show later, this not only reduces the $\hat{\tau}$ matter density contrast but also retards the growth of the $\hat{\Omega}$ fluctuations. The slower growth of the $\hat{\Omega}$ structure results in a stronger suppression of the matter power spectrum at low redshift, as is shown in the blue curve of Fig.~\ref{fig:cartoon}. 

Here we describe the way in which the acoustic oscillations experienced by the $\hat{\tau}$-DR system act to suppress $\sigma_8$. A more detailed study is presented in Ref.~\cite{Chacko:2016kgg}. We employ the general formalism of Ma and Bertschinger \cite{Ma:1995ey} for scalar perturbations in the conformal Newtonian gauge. 
Working in momentum space, we express the coupled evolution equations in terms of the comoving wavenumber $k$ and conformal time derivative $\dot{\phantom{x}}=1/d\tau$. 
Then the evolution of the over-density of $\hat{\Omega}$ can be described by the linear equations
\begin{equation}\label{eq:CDM}
\dot{\delta}_{\hat{\Omega}}=-\dot{\theta}_{\hat{\Omega}}+3\dot{\phi},\quad\dot{\theta}_{\hat{\Omega}}=\frac{\dot{a}}{a}\theta_{\hat{\Omega}}+k^2\psi.
\end{equation}
Here the over-density, $\delta_s\equiv \delta\rho_s/\bar{\rho}_s$, parametrizes the density perturbation relative to the average density of matter or radiation. Further, the parameter  
$\theta_s=\partial_j v^j_s$ is the divergence of a comoving 3-velocity, which modifies the density perturbation by having particles move out of the overdense region. In addition, $\psi$ and $\phi$ are the metric perturbations in the conformal Newtonian gauge $ds^2=a(\tau)^2[-(1+2\psi)d\tau^2+(1-2\phi)dx^idx^j]$. Since the dark radiation is tightly coupled, we take $\phi=\psi$ in the equations and ignore the minor correction from SM neutrinos. We note that Eq.~(\ref{eq:CDM}) coincides with the corresponding evolution equation for the density perturbation of standard cold dark matter particles for a given metric perturbation $\phi$, where $\phi$ evolves according to the Einstein equation 
\begin{equation}
k^2\phi + 3\frac{\dot{a}}{a} \!\left( \dot{\phi} + \frac{\dot{a}}{a} \psi \right)\!
= -4\pi G a^2 \sum_s\rho_s\delta_s 
\,.\label{eq:evolution:psi}
\end{equation}
The source of the gravity perturbation is dominated by the matter or radiation density with the largest $\rho_s\delta_s$ contribution on the RHS. 

The evolution equations for the interacting dark matter component $\hat{\tau}$ are given by
\begin{equation}\label{eq:IntDM}
\dot{\delta}_{\hat{\tau}}=-\dot{\theta}_{\hat{\tau}}+3\dot{\phi},\quad\dot{\theta}_{\hat{\tau}}=\frac{\dot{a}}{a}\theta_{\hat{\tau}}+k^2\psi+a\Gamma(\theta_{\text{DR}}-\theta_{\hat{\tau}}),
\end{equation}
where $\Gamma\equiv\langle p^2_{\hat{\tau}}\rangle^{-1}d\langle\delta p^2_{\hat{\tau}}\rangle/dt$ is the thermal averaged momentum transfer rate experienced 
by a $\hat{\tau}$ particle as it travels through the dark fluid. Note that here $t$ is the Minkowski time, since the rate $\Gamma$ is a microscopic quantity independent of the cosmological expansion. In the twin sector, this scattering rate is given by \cite{Buen-Abad:2015ova}
\begin{equation}
\Gamma=\frac{8\pi}{9}\alpha_{\hat{L}}^2\ln\alpha_{\hat{L}}^{-1}\frac{\hat{T}^2}{m_{\hat{\tau}}},
\end{equation}
where $\hat{T}$ is the temperature of $\hat{\tau}$. In the tightly coupled limit, $\hat{T}$ equals the temperature of dark radiation.

The tight-coupling approximation is valid as long as the interaction rate $\Gamma$ is comparable to or exceeds the Hubble rate during structure formation. We focus on the case for which $\Gamma$ is significantly larger than the Hubble expansion rate, which enables us to gain an analytical understanding of the oscillation physics. This is easily achieved provided that the U$(1)_{\hat{L}}$ coupling satisfies
\begin{equation}
\alpha_{\hat{L}}
\gg 10^{-8}\sqrt{\frac{m_{\hat{\tau}}}{1\,\text{GeV}} } \, \biggl( \frac{T_0}{\hat{T}} \biggr)
\,,\label{eq:Gamma>>H0}
\end{equation} 
where $T_0$ is the photon temperature today and $\hat{T}\simeq 0.4\,T_0$. 
In this tightly coupled limit, we have that $a\Gamma\gg aH\simeq\tau^{-1}$, which implies that the scattering term in Eq.~(\ref{eq:IntDM}) dominates the $\hat{\tau}$ evolution. 
The consequence is that $\theta_{\hat{\tau}}=\theta_{\text{DR}}$, which authorizes us to combine the evolution equations of the dark matter component $\hat{\tau}$ and the DR as follows:
\begin{eqnarray}
\dot{\delta}_{\text{DR}} &=& -\frac43\theta_{\text{DR}} + 4\dot{\phi},\label{eq:evolution:DR}
\\
\dot{\theta}_{\text{DR}} &=& k^2 \Bigl( \frac{\delta_{\text{DR}}}{4} + \psi \Bigr) + \frac34 \frac{\bar{\rho}_{\hat{\tau}}}{\bar{\rho}_{\text{DR}}} a\Gamma (\theta_2 - \theta_{\text{DR}}),
\nonumber
\end{eqnarray}
to obtain ($\hat{R}\equiv 3\rho_{\hat{\tau}}/4\rho_{\text{DR}}$)
\begin{equation}\label{eq:oscillation}
\quad\ddot{\delta}_{\hat{\tau}}+\frac{\dot{a}}{a}\frac{\hat{R}}{1+\hat{R}}\dot{\delta}_{\hat{\tau}}+\frac{k^2}{3(1+\hat{R})}\delta_{\hat{\tau}}\simeq-k^2\phi.
\end{equation}

The resulting evolution equations indicate that in the tight-coupling limit, the $\hat{\tau}-$DR system behaves like a single coupled fluid. Here we neglect terms containing higher-order conformal time derivatives of $\psi$ in order to focus on modes well within the horizon, for which $k\tau\gg 1$. We find that the evolution of the over-density $\delta_{\hat{\tau}}$ is similar to that of the SM baryon. Let us now consider varying $\hat{R}$. In the regime $\hat{R}\ll 1$, the $\hat{\tau}$-DR \emph{fluid} is relativistic. In this limit, the first two terms in Eq.~(\ref{eq:oscillation}) generate an acoustic oscillation of $\delta_{\hat{\tau}}$ without building up the density perturbation. This has the effect of delaying the growth of dark matter structure when the mode enters horizon, as is shown by the PAcDM (blue) and QuAcDM (orange) curves in Fig.~\ref{fig:cartoon}. It is only when the dark radiation cools down and one enters the $\hat{R}\gg 1$ regime that the power spectrum begins to grow monotonically.

This delay of the $\hat{\tau}$ structure formation results in $\delta_{\hat{\tau}}\ll \delta_{\hat{\Omega}}$ 
upon entering the matter domination era. 
Inserting the density ratio $r$ defined in Eq.~(\ref{eq:masses}), we can then express the total dark matter density perturbation as
\begin{equation}
\frac{\delta\rho_{\text{DM}}}{\bar{\rho}_{\text{DM}}}=[(1-r)\delta_{\hat{\Omega}}+r\delta_{\hat{\tau}}]\simeq(1-r)\delta_{\hat{\Omega}}.
\end{equation}
For modes well within the horizon, $\tau k\gg1$, so that Eq.~(\ref{eq:evolution:psi}) simplifies to
\begin{equation}
k^2 \phi\simeq -4\pi a^2 G \bar{\rho}_{\text{DM}} \cdot (1-r) \delta_{\hat{\Omega}}= -\frac{6}{\tau^2}\,(1-r)\,\delta_{\hat{\Omega}},
\end{equation}
where we have applied the Friedman equation and noted that the scale factor $a$ is proportional to $\tau^2$ during matter domination. Canceling the 
$\theta_{\hat{\Omega}}$ in Eq.~(\ref{eq:CDM}) and inserting the above metric perturbation, we then have for the dominant dark matter component ($\eta\equiv k\tau$, $\phantom{x}'\equiv 1/d\eta$)
\begin{equation}
\delta''_{\hat{\Omega}}+\frac{2}{\eta}\delta'_{\hat{\Omega}}\simeq\frac{6(1-r)}{\eta^2}\delta_{\hat{\Omega}}.
\end{equation}
Hence, the growth of the dark matter density perturbation obeys a reduced power law
\begin{equation}
\delta_{\hat{\Omega}}\propto\left(\frac{a}{a_{\text{eq}}}\right)^{1-0.6r+\mathcal{O}(r^2)}.
\end{equation}

Further, since the power spectrum $P$ of the dark matter density perturbation is proportional to the square of the over-density, $\delta_{\text{DM}}^2$, we find that the ratio of the power spectrum with and without the interacting component $\hat{\tau}$ is given by
\begin{equation}\label{eq:psratio}
\frac{P(r)}{P(0)}\simeq\frac{(1-r)^2\delta^2_{\hat{\Omega}}(r)}{\delta^2_{\hat{\Omega}}(0)}\simeq(1-2r)\left(\frac{a}{a_{\text{md}}}\right)^{-1.2r},
\end{equation}
where $a_{\text{md}}\simeq 10^{-3}$ is the scale factor at which matter dominates the source term of the Einstein equation in Eq.~(\ref{eq:evolution:psi}). In order to eliminate the $\sigma_8$ discrepancy through the reduction of the density perturbation by $\simeq 10\%$, we need to suppress the matter power spectrum in Eq.~(\ref{eq:psratio}) by $\simeq 20\%$. This requires $r\simeq 2-3\%$ and is the reason for the benchmark value in Eq.~\ref{eq:benchmark}.

To obtain a more precise result, we determine the size of the over-density $\delta_{\text{DM}}$ by numerically solving equations Eq.~(\ref{eq:CDM})-(\ref{eq:IntDM}) and (\ref{eq:evolution:DR}), where we 
choose a U$(1)_{\hat{L}}$ coupling that satisfies Eq.~(\ref{eq:Gamma>>H0}). We also incorporate the evolution equations for the SM photon and baryon by making the replacements DR$\to\gamma$ and $\hat{\tau}\to$ SM baryon (B), respectively. For the modes that enter the horizon during radiation (matter) domination, the initial conditions for solving the coupled system are given by
\begin{equation}
\delta_{\gamma, \text{DR}}^{r(m)} = \frac43 \delta_{\hat{\Omega},\hat{\tau},\text{B}} = \xi_1^{r(m)} \psi
\,,\quad
\theta_{\hat{\Omega},\hat{\tau},\text{DR}, \text{B}, \gamma} =  \xi_2^{r(m)} k^2 \tau \psi 
\,,\label{eq:evolution:initial}
\end{equation}
with the values $\xi_1^r = -2$, $\xi_1^m =-\frac83$, $\xi_2^r =  \frac{1}{2}$, and $\xi_2^m=  \frac{1}{3}$. We make the following parameter choices when solving this coupled system of evolution equations: 
$h=0.68$, $\Omega_{\gamma} h^2 = 2.47 \times 10^{-5}$, $\Omega_{\Lambda} h^2 = 0.69$, $\Omega_{b} h^2 = 2.2\times 10^{-2}$ and $\Omega_{\nu}=0.69\, \Omega_{\gamma}$ \cite{Ade:2015xua}. We note that the dark energy density $\Omega_\Lambda$ has only a small effect so that its precise value is not important for our purpose here. We choose ${N_{\text{eff}}^\text{scatt}} = 0.46$, assuming the presence of the anomaly compensator $\hat{\nu}'$, and a slightly larger value of $\Omega_{\text{DM}}h^2 = 0.13$ in order to keep the redshift at matter-radiation equality unchanged. This allows us to compare our matter power spectrum to that of a conventional single-component dark matter model without any dark radiation. 

We find that the choice of $r=2.5\%$ leads to an $8\%$ suppression of the density perturbation around the scale $k \sim 0.2 h$ Mpc$^{-1}$ as compared to $\Lambda$CDM, thereby solving the $\sigma_8$ problem. It should be noted that this corresponds to a $\simeq 23\%$ suppression when compared to the $r=0$ case with the same amount of dark radiation, as displayed in the left panel of Fig.~\ref{fig:psratio}. 
In the same plot, we also show the ratio of the power spectrum, $P(r)/P(0)$, during the CMB time with $a\sim 10^{-3}$ (dashed curve). Due to the redshift dependence in Eq.~(\ref{eq:psratio}), the suppression of the matter power spectrum is smaller at the earlier CMB time, and hence the correction to the metric perturbation is minor at this time. This feature allows the model to flexibly accommodate bounds from the CMB temperature spectrum and CMB lensing, as is discussed in Ref.~\cite{Chacko:2016kgg}.

If we consider an alternative scenario, in which both $\hat{\Omega}$ and $\hat{\tau}$ scatter with the dark radiation through a gauged twin B$-$L symmetry as a realization of the QuAcDM framework, we can obtain the same suppression by requiring $\alpha_{\hat{B}-\hat{L}}\sim 10^{-9.8}$. Distinct from the PAcDM setup, the QuAcDM scenario gives comparable suppressions to the matter density perturbation between today and the CMB time. As is shown in Fig.~\ref{fig:cartoon}, the different corrections to the power spectrum provide a way to differentiate between these dark sector scenarios through the CMB observation.
\section{Small-scale Structure: Twin Baryon with a Self-interaction}\label{sec:SIDM}

In this section, we study the dark matter self-interaction through twin photon exchange (Fig.~\ref{fig:scattering}) and make a connection to dark matter halo structures. Our twin sector contains the self-interacting dark matter particles 
$(\hat{\Omega},\,\hat{\tau})$ and a dark fluid scattering with $\hat{\tau}$, which gives rise to complicated dynamics for halo formation. A detailed N-body simulation of the halo structure is beyond the scope of this work. In the present analysis, we only focus on the dominant dark matter component $\hat{\Omega}$, which contains $\simeq 98\%$ of the dark matter density, and comment on the possible correction to the result from the presence of $\hat{\tau}$ and the dark fluid.

The halo structure formation depends on the average time scale of dark matter scattering $\langle n\sigma v\rangle^{-1}$, where $n=\rho_c\Omega_{DM} /m_{\text{DM}}$ gives the number density, and the various relevant effects are determined by the cross section mass ratio $\sigma/m_{\text{DM}}$. In order to solve the mass deficit problem from dwarf galaxy to galaxy cluster scales, we require that $\sigma/m_{\text{DM}}$ be $\sim 1$ cm$^2/$g for dwarf galaxies and $\sim 0.1$ cm$^2/$g for galaxy clusters. One way to achieve this is to introduce a velocity-dependent dark matter self-scattering process with the mediator mass comparable to or lighter than the momentum exchange of the dark matter particles. Since the average collision velocity between dark matter particles in dwarf galaxies is about an order of magnitude smaller than the corresponding value for the galaxy clusters, the nonperturbative effects in the scattering cross section, enhanced by a low dark matter velocity, can help to reconcile the required $\sigma/m_{\text{DM}}$ for different objects. 

In this work, we estimate the size of $\sigma/m_{\text{DM}}$ by applying standard partial wave methods discussed in Ref.~\cite{Tulin:2013teo} to a range of twin photon masses and couplings. We focus on scattering outside of the Born regime, demanding that $\hat{\alpha}m_{\hat{\Omega}}/m_{\hat{\gamma}}\gsim 1$, so that the nonperturbative effects of nonrelativistic scattering become important. 
Since the dark matter density contribution $\Omega_{\hat{\Omega}}$ arises from the twin baryon asymmetry, the dark matter particles are scattered by a repulsive potential $V(r)=\hat{\alpha}e^{-m_{\hat{\gamma}}r}/r$ with fine structure constant $\hat{\alpha}$ in the twin sector, and the transfer cross section of dark matter scattering can then be expressed as ($\beta\equiv2\hat{\alpha}m_{\hat{\gamma}}/m_{\hat{\Omega}}v^2$) \cite{Tulin:2013teo}
\begin{equation}\label{eq:sidm1}
\sigma_T=\begin{cases}\frac{2\pi}{m_{\hat{\gamma}}^2}\beta^2\ln(1+\beta^{-2}) & \beta\lsim 1 \\ \frac{\pi}{m_{\hat{\gamma}}^2}(\ln 2\beta-\ln\ln2\beta)^2 & \beta\gsim 1 \end{cases},
\end{equation}
in the classical limit $m_{\hat{\Omega}}v/m_{\hat{\gamma}}\gg 1$. For $m_{\hat{\Omega}}v/m_{\hat{\gamma}}\lsim 1$, a good approximation is obtained by
\begin{equation}\label{eq:sidm2}
\sigma_T=\frac{16\pi}{m_{\hat{\Omega}}^2v^2}\sin^2\delta_0,\quad\delta_0=\text{arg}\left(\frac{i\Gamma(\frac{im_{\hat{\Omega}}v}{\kappa m_{\hat{\gamma}}})}{\Gamma(\lambda_+)\Gamma(\lambda_-)}\right),
\end{equation}
with $\kappa\approx 1.6$ and 
\begin{equation}
\lambda_{\pm}\equiv 1+\frac{im_{\hat{\Omega}}v}{2\kappa m_{\hat{\gamma}}}\pm i\sqrt{\frac{\hat{\alpha}m_{\hat{\Omega}}}{\kappa m_{\hat{\gamma}}}+\frac{m_{\hat{\Omega}}^2v^2}{4\kappa^2m_{\hat{\gamma}}^2}}\,.
\end{equation}
Since the number density of the subcomponent dark matter $\hat{\tau}$ is comparable to that of the dominant dark matter 
$\hat{\Omega}$, the chance of having a $\hat{\Omega}$ particle scatter with $\hat{\tau}$ is therefore comparable to the $\hat{\Omega}$ self-scattering. However, given that $m_{\hat{\Omega}}\gg m_{\hat{\tau}}$, the momentum transfer from the $\hat{\Omega}$-$\hat{\tau}$ scattering is accordingly much smaller than that from the $\hat{\Omega}$ self-scattering. It is then reasonable to consider only the $\hat{\Omega}$ scattering to a good approximation.

Our study focuses on dark matter halo structure anomalies in dwarf galaxies, low surface brightness galaxies (LSBs), and galaxy clusters. Instead of fitting the result for each of \text{these} objects, we approximate the results for the ratio of the cross section to the dark matter mass in Ref.~\cite{Kaplinghat:2015aga} in terms of various ranges of this ratio. For dwarf galaxies, we take 
$\sigma /m_{\text{DM}}=0.5-5$ cm$^2/$g and $v=60$ km$/$s. Next, for LSB galaxies, we assume $\sigma/m_{\text{DM}}=0.5-5$ cm$^2/$g and $v=100$ km$/$s. Further, for galaxy clusters, we take $\sigma/m_{\text{DM}}=0.05-0.5$ cm$^2/$g and $v=1200$ km$/$s. 
When studying the cross section, we also consider bounds from the 
ensemble of merging clusters of $\sigma /m_{\text{DM}}<0.47$ cm$^2/$g at $95\%$ CL \cite{Harvey:2015hha} at a collision velocity of $v=900$ km$/$s.
\begin{figure}
\centering
\includegraphics[width=7.4cm]{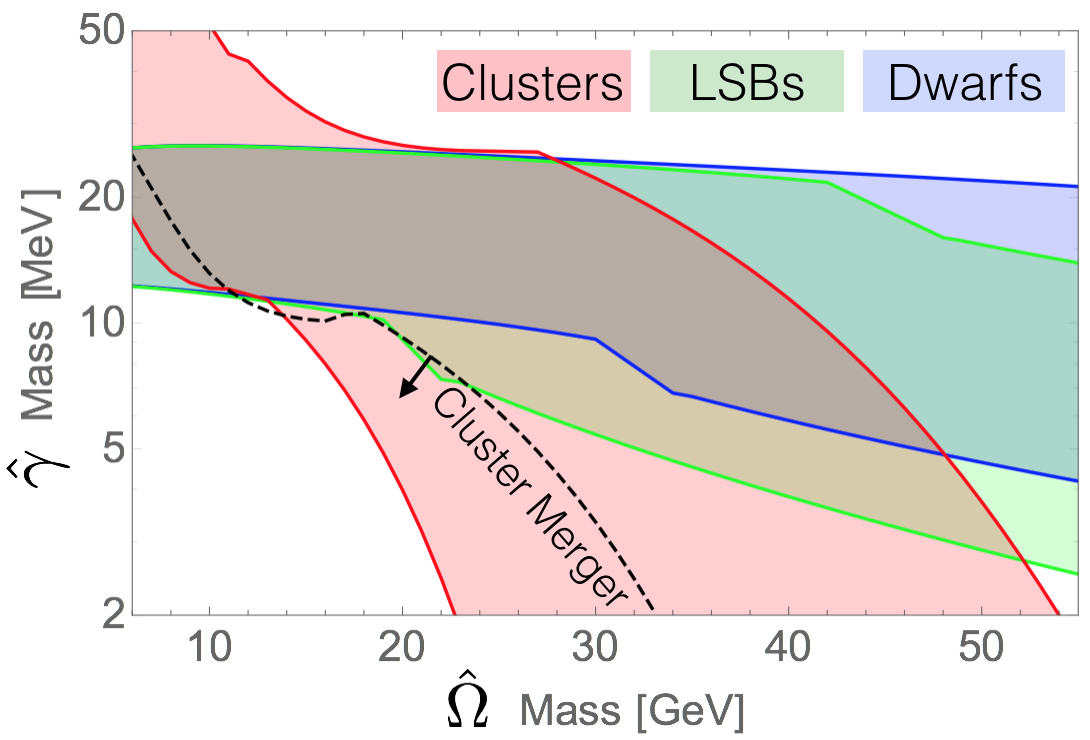}\quad\vspace{2em}\includegraphics[width=7.7cm]{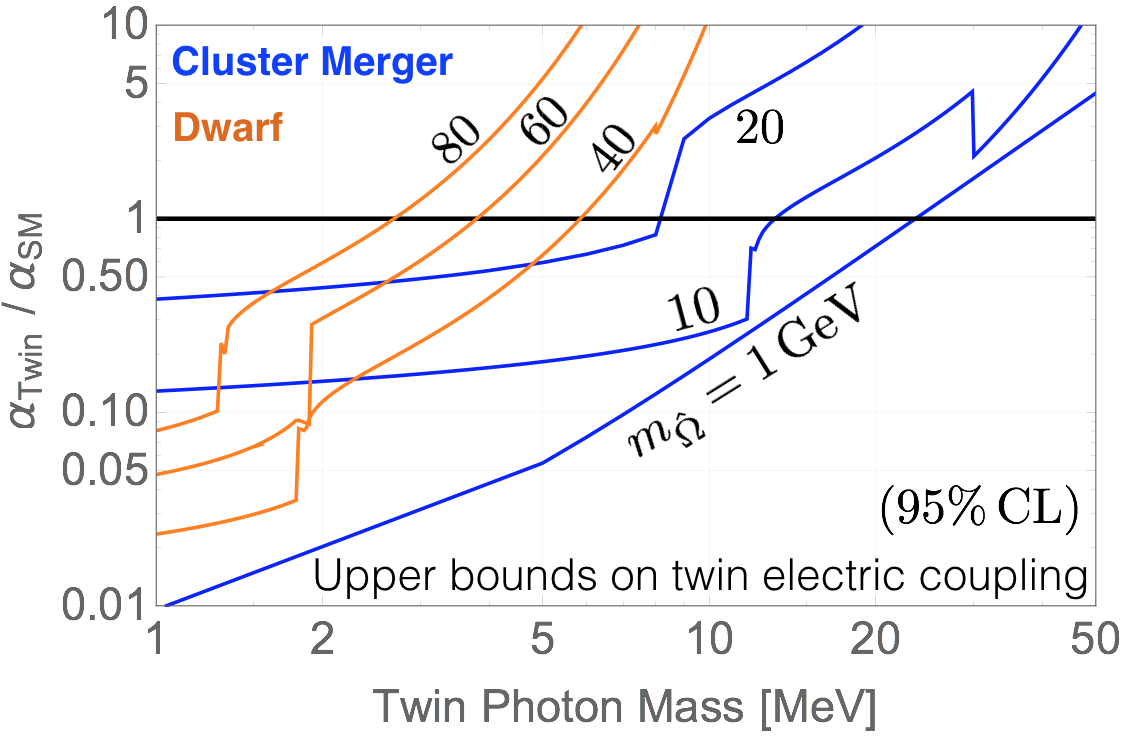}
\caption{Left: SIDM parameter space when considering only the dominant dark matter component $\hat{\Omega}$ 
and a $\mathbb{Z}_2$-symmetric twin photon coupling $\hat{\alpha}=\alpha$. See Sec.~\ref{sec:SIDM} for the choice of cross section and velocity values. The dashed curve shows a lower bound on the dark matter mass from the cluster merger constraint. The overlap area (gray) among different allowed regions gives solutions to the mass deficit problem for the three types of galactic objects. Right: Upper bounds on the twin electromagnetic coupling from the cluster merger and dwarf halo constraints. The kinks of the curves in both plots correspond to the transition points between different  analytical approximation regimes in Eq.~(\ref{eq:sidm1}) and (\ref{eq:sidm2}) for the cross section calculation.For example, in the $m_{\hat{\Omega}}=10$ GeV cluster merger curve, the first kink from the left corresponds to $\beta\simeq1$, and the second kink corresponds to $m_{\hat{\Omega}}v/m_{\hat{\gamma}}\simeq1$.}\label{fig:SIDM}
\end{figure}

In the upper panel of Fig.~\ref{fig:SIDM}, we show the allowed sizes of $m_{\hat{\gamma}}$ and $m_{\hat{\Omega}}$ required to solve the mass deficit problem in various galactic objects. The plot assumes $\mathbb{Z}_2$-symmetric electromagnetic couplings $\hat{\alpha}=\alpha$ between the twin and SM sectors. It turns out that if we fix the mass parameters to be in the range $10\lsim m_{\hat{\Omega}}\lsim 40$ GeV and $10\lsim m_{\hat{\gamma}}\lsim 20$ MeV, the self-interaction of the twin baryon can provide a plausible solution to the small-scale structure problem. 

In addition to explaining the anomaly, an analysis of the effect of dark matter self-scattering on halo formation also sets a bound on the twin photon interaction. When we consider a twin sector that contains stable charged baryons carrying a SM-like baryon asymmetry, the self-coupling of the baryons is subject to an upper bound that ensures that the self-scattering does not violate the small-scale structure constraints. In the lower panel of Fig.~\ref{fig:SIDM}, we investigate the upper bounds on the twin photon coupling by applying constraints from the merging cluster with $\sigma /m_{\hat{\Omega}}<0.47$ cm$^2/$g and the shape of dwarf halos with $\sigma /m_{\hat{\Omega}}<5$ cm$^2/$g. If we suppose that the twin baryon dominates the dark matter density,  there then needs to be a breaking of the mirror symmetry either through a nonzero $m_{\hat{\gamma}}$ or a smaller twin electric coupling.

When we consider the effect of the presence of $\hat{\tau}$ and the dark fluid, then no matter whether $\hat{\tau}$ contributes to a core- or a cusp-like density profile, the $2.5\%$ dark matter density does not yield observable signatures in the current measurements of halo structure. 
One potential application of the $\hat{\tau}$-dark fluid scattering is the following:
If the dark fluid is able to cool down the $\hat{\tau}$ particles enough such that the subcomponent dark matter falls into the galactic center, this mechanism would provide a possible explanation of the origin of supermassive blackholes \cite{Pollack:2014rja}. 
When nonlinear halo formation sets in around a redshift of $z=10-20$ \cite{Planelles:2014zaa}, the dark fluid is much colder than the virial temperature of $\hat{\tau}$, given by $\simeq (m_{\hat{\tau}}/1\,\text{GeV})$ keV. 
So, if the fluid is able to rapidly transport heat outside the halo, the dark matter $\hat{\tau}$ can accordingly undergo efficient cooling and collapse into a black hole. However, since the free streaming length of the dark fluid is in fact very short, 
expected to be only $\sim 10$ m for a twin coupling of size $\alpha_{\hat{L}}\simeq 10^{-2}$ and a fluid temperature of about $\hat{T}\sim10^{-4}$ eV, a dark fluid particle makes a random walk across a distance $\sim 10^{-3}$ pc until today \cite{Curtin:2014afa}, which is negligible relative to a $\sim 10$ kpc-size halo.  Hence, we conjecture that the dark fluid does not dissipate heat efficiently through diffusion and expect that a better cooling mechanism such as convection is required to form the black hole.

\section{Solutions from The Twin Hydrogen DM}\label{sec:twinH}
The extended Fraternal Twin Higgs model described above contains two questionable assumptions of the $\mathbb{Z}_2$ breaking. 
The first one is that in order to accommodate the mass deficit problem, it is necessary to break the twin electromagnetism $U(1)_L$ at the MeV scale, thus introducing an additional scale which is not associated with any other mass scales in the model. 
The second assumption is that if we choose to gauge the U$(1)_L$ symmetry, we are compelled to include anomaly compensators in the SM. These suffer from strong experimental constraints, and an additional $\mathbb{Z}_2$ breaking is required to lift the compensator mass.

Rather than adding extra layers of the model, here we present an alternative solution to the structure problems that does not require the twin U$(1)_{EM}$ breaking and the introduction of additional fermions in the TH model. We will see below that in this scenario, the required $\mathbb{Z}_2$ breaking will be a $\simeq 60\%$ deviation between the SM and twin electric couplings and that the solution will feature different Yukawa couplings of the light fermions.

Let us first describe the main idea of the model. When the temperature of the twin sector drops below the twin confinement scale, 
we assume that the twin sector contains the following spectrum: the twin proton $\hat{p}^+$, twin lepton $\hat{\ell}^-$, light twin neutrinos $\hat{\nu},\,\hat{N}$, as well as massless gauge bosons $\hat{\gamma},\,\hat{\gamma}_{B-L}$. Instead of gauging the twin lepton number, here we gauge the twin U$(1)_{B-L}$ that is anomaly-free assuming the presence of right-handed neutrinos. In this scenario, we can break the SM U$(1)_{B-L}$ above the EW scale through the same $\mathbb{Z}_2$ breaking as in the Higgs potential. For example, upon getting a VEV, the $\mathbb{Z}_2$-odd scalar $\eta$ in the $\mathbb{Z}_2$-symmetric potential $\mathcal{L}\subset -\mu\eta(|H_A|^2-|H_B|^2)+\mu'\eta(|\phi_A|^2-|\phi_B|^2)$ can induce the breaking of SM U$(1)_{B-L}$ through the $B-L$ charged scalar $\phi$ and split the EWSB scales by $\sim\sqrt{\mu\langle\eta\rangle}$ in the two sectors. In order to achieve a successful PAcDM scenario, the U$(1)_{B-L}$ coupling required to resolve the LSS problem can be as small as $\alpha_{B-L} \sim10^{-8}$, and existing constraints from the $Z'$ search only cover the TeV scale $\gamma_{B-L}$ with $\alpha_{B-L}\gsim 10^{-4}$ \cite{Okada:2016gsh}.

Twin hydrogen $\hat{H}$ starts to form when the twin temperature drops below the binding energy between $(\hat{p}^+\,\hat{\ell}^-)$. Following recombination, a small fraction of the twin particles remains ionized. Since $\hat{H}$ is neutral under both the U$(1)_{EM}$ and U$(1)_{B-L}$ symmetries, it behaves as a cold DM particle during the structure formation. Meanwhile, the few ionized twin particles $(\hat{p}^+,\,\hat{\ell}^-)$ scatter with the twin neutrinos via the t-channel U$(1)_{B-L}$ process. This process realizes the PAcDM framework in this scenario, furnishing the mechanism which suppresses $\sigma_8$ and enhances the Hubble value due to the presence of additional twin radiation.

During the DM halo formation, the virial temperature of the dark plasma is lower than the binding energy. Hence, the twin hydrogen $\hat{H}$ remains stable and constitutes the dominant DM component inside halos. Characterized by an extended but finite size, the $\hat{H}$ atom furnishes a good SIDM candidate if its geometric size is around the barn scale. The reason stems from the property that the scattering between two $\hat{H}$'s contains both elastic and inelastic processes. The elastic process comes from the collision between two atoms, which transfers energy from one atom to the other and acts to keep DM thermalized. Meanwhile, the inelastic process comes from the hyperfine splitting between $\hat{\ell}$ and $\hat{p}$. In the inelastic case, when $\hat{\ell}$ absorbs part of the collisional energy into the excited state, the subsequent decay into the ground state releases DM energy into soft $\hat{\gamma}$. This cooling process is important when $\hat{H}$ carries a kinetic energy comparable to the hyperfine splitting, in which case the scattering process introduces an additional velocity dependence in the scattering cross section. A numerical study of the scattering cross section has been performed in \cite{Boddy:2016bbu}. From this, it emerges that the general trend is that the cross sections of dwarf halo particles with lower DM velocities tend to be larger than those of particles in cluster halos with higher velocities, which feature provides the correct behavior required to solve mass deficit problems in both cases. 

We rely on the results in Ref.~\cite{CyrRacine:2012fz,Boddy:2016bbu} to determine reasonable parameters in the twin sector for our purposes. It turns out that in order to solve both the large- and small- scale structure problems simultaneously, we need the twin masses $m_{\hat{p}}\approx 20$ GeV, $m_{\hat{\ell}}\approx 3$ GeV, the twin electric coupling $\hat{\alpha}\approx 0.02$, and the twin U$(1)_{B-L}$ coupling $\hat{\alpha}_{B-L}\gsim 10^{-9}$. The last bound is necessary to ensure that the t-channel $\hat{p}\,\hat{\nu}\to\hat{p}\,\hat{\nu}$ scattering is effective during structure formation.

We next discuss some details of the parameters. First, the fraction of ionized atoms can be approximated as \cite{CyrRacine:2012fz}
\begin{equation}
\chi_e\sim2\times 10^{-16}\frac{\xi}{\alpha_d^6}\left(\frac{0.11}{\Omega_{\text{DM}}h^2}\right)\left(\frac{m_{\text{H}}}{\text{GeV}}\right)\left(\frac{B_{\text{H}}}{\text{keV}}\right).
\end{equation}
Here $\xi\simeq 0.5$ is the ratio between the Twin and SM temperature today when the twin radiation contributes $\Delta N_{eff}\simeq 0.4$ for solving the $H_0$ problem. $B_{\text{H}}=\alpha_d^2\,\mu_{\text{H}}/2$ is the binding energy of the dark hydrogen atom, and $\mu_{\text{H}}$ is the reduced mass of the $\hat{p}\,\hat{\ell}$ system. Upon choosing $\hat{\alpha}=0.02$, $m_{\hat{p}}=20$ GeV, and $m_{\hat{\ell}}=3$ GeV, we find that the resulting fraction is $\chi_e\simeq 2.5\%$. As discussed in Sec.~3, this value results in the appropriate amount of oscillating DM necessary to solve the $\sigma_8$ problem. 

For simplicity, we assume the U$(1)_{B-L}$ coupling to be smaller than the electric coupling, $\hat{\alpha}_{B-L}<\hat{\alpha}$, so that the twin EM dominates the binding force. With the above choices, the binding energy is $B_H\approx 400$ keV, which significantly exceeds the virial temperature $\simeq 2$ keV of the twin atom during galaxy formation\footnote{The virial temperature is given by $T_{\text{vir}}\approx 0.9\,\text{keV}\frac{M}{M_{\text{DM}}^{\text{gal}}}\frac{\mu}{10\,\text{GeV}}\frac{110\,\text{kpc}}{R_{\text{vir}}}$, where $M$ represents the mass of the virial cluster, $M_{\text{DM}}^{\text{gal}}=10^{12}M_{\odot}$ is the mass of DM in the Milky Way galaxy, and $\mu$ is the average mass of a dark plasma particle (in our case this is $\mu=\rho_{\hat{p}}/n_{\hat{p}}=20$ GeV) \cite{Randall:2007ph}.}. Moreover, for the size of $\hat{\alpha}$ we choose, the rate of collisional ionization between $\hat{H}$ atoms in cluster halos is always smaller than the cluster lifetime \cite{Boddy:2016bbu}. Hence, the twin atom remains in the ground state during halo formation.

If we consider the case where the $\hat{H}$ atom is composed of a spin-$\frac{1}{2}$ $\hat{p}$ and a spin-$\frac{1}{2}$ $\hat{\ell}$, the hyperfine splitting for the chosen mass and coupling parameters is $\sim 100$ eV. The value is above the $\hat{\ell}$ energy due to the virial velocity inside dwarf galaxies but below that inside galaxy clusters. Therefore, the hyperfine splitting plays a more important role in DM scattering inside galaxy clusters than dwarf galaxies. From the numerical study in \cite{Boddy:2016bbu}, such an energy splitting generates different DM thermalization effects at dwarf galaxies and galaxy clusters, which feature enables one to successfully solve the mass deficit problem in both systems.  

Meanwhile, in the single-generation model, the lightest twin baryon $(\hat{b}\hat{b}\hat{b})^-$ is a spin-$\frac{3}{2}$ particle, and a numerical study of the hyperfine splitting between a spin-$\frac{3}{2}$ nucleus and a spin-$\frac{1}{2}$ lepton in the $\hat{H}$ scattering is more involved and is beyond the scope of this work. In order to give a viable example of the model, we simply adapt the result in \cite{Boddy:2016bbu} by taking the nucleus to be a spin-$\frac{1}{2}$ particle. We then assume two generations of twin fermions and 
take $(\hat{b},\,\hat{s})$ to be the lightest twin quarks. In this case, $(\hat{b},\,\hat{s})$ quarks remain stable and form protons, $\hat{p}^-=(\hat{b}\hat{b}\hat{s})^-$ or $(\hat{b}\hat{s}\hat{s})^-$, while the twin neutron is absent due to a fast $\hat{c}\to\hat{s}\hat{\mu}\bar{\hat{\nu}}$ decay. With two generations of quarks, RG running gives the twin confinement scale $\hat{\Lambda}\simeq 270$ MeV.  Here we assume that $m_{\hat{b}}\simeq m_{\hat{s}}$. Although this choice introduces additional $\mathbb{Z}_2$ breaking, it does not ruin Higgs tuning owing to the smallness of the Yukawa couplings for the $\hat{b}$ and $\hat{s}$ quarks and constitutes a better alternative to the introduction of anomaly compensators.
Then, approximating the twin proton mass as $m_{\hat{p}}\simeq 3(m_{\hat{s}}+\hat{\Lambda})$, we find that the lightest twin quarks feature a mass of $6.5$ GeV. Twin muon carries a mass of $3$ GeV and is combined with twin proton to form the twin hydrogen. In this case, the lightest twin hadron is the scalar glueball $m_{\hat{G}_{0^{++}}}\simeq 2$ GeV, which decays promptly into SM muons (with lifetime $\sim 10^{-6}$ sec) during the twin confinement.

In this model, we find that when the SM and twin sector decouple around the GeV scale, the light degrees of freedom from the dark radiation $\hat{\nu}_{\tau,\mu},\,\hat{N}_{\tau,\mu},\,\hat{\gamma},\,\hat{\gamma}_{B-L}$ contribute a $\Delta N_{\text{eff}}=0.4$, which is
the required value for solving the Hubble problem. We thus see that this Twin Hydrogen scenario successfully resolves both the large- and 
small-scale structure problems without introducing additional mass scales or anomaly compensators.

\section{Discussion and Conclusions}\label{sec:conclusion}

In this work, we take the viewpoint that the Hierarchy problem and the large- and small-scale structure anomalies
are all indicative of the existence of a dark sector that extends beyond the $\Lambda$CDM paradigm. We investigate potential solutions to these problems in the context of an extension of the Fraternal Twin Higgs model, which contains only the heavier-generation partners of SM fermions and a massless gauge boson that gives the DM and twin neutrino scattering. We first discuss the $\hat{\Omega}$-$\hat{\tau}$ scenario, which assumes a SM-like baryogenesis, the twin baryon $\hat{\Omega}\sim(\hat{b}\hat{b}\hat{b})$ and the twin tau $\hat{\tau}$ become metastable dark matter particles due to twin baryon number and an approximate twin U$(1)_{\text{em}}$ symmetry. 
Through the exchange of a $\sim 10$ MeV-scale twin photon, these dark matter particles have a self-interaction cross section that successfully resolves the mass deficit problem on all scales, from the dwarf galaxies to the galaxy clusters.
In the specific implementation of the PAcDM framework that we consider, the gauged U$(1)_{\hat{L}}$ force acts to suitably damp the dark matter power spectrum, supposing it remains effective at the beginning of matter domination. In particular, if $m_{\hat{\tau}}\simeq 2.5\%\,m_{\hat{\Omega}}$, such a damping can indeed reduce the size of $\sigma_8$ by $\simeq 8\%$, reconciling the $\sigma_8$ discrepancy. 
Moreover, the overall $\Delta N_{\text{eff}}^{\text{scatt}}$, which receives contributions from the tightly coupled fluid in the twin sector, including the massless neutrinos, U$(1)_{\hat{L}}$ gauge boson, and its anomaly compensators, is able to flexibly enhance the size of $H_0$. Favored by a weaker CMB constraint on a tightly coupled fluid, this model can efficiently reconcile the tension between the $H_0$ results from the CMB and local measurements. 

We also discuss the scenario of the twin hydrogen DM. In this case, $\simeq 2.5\%$ twin protons remain ionized, and their scattering to twin neutrinos through a twin U$(1)_{B-L}$ force damps the matter power spectrum realizing the PAcDM framework. The twin photon remains massless in this case, and the scattering between two twin hydrogens contains the required velocity dependence necessary to successfully resolve the mass deficit problem from dwarf galaxy to galaxy cluster scales.

Our study is based on the Fraternal Twin Higgs model, which contains a smaller number of neutrinos than the full three-generation case and hence is able to more easily satisfy the $\Delta N_{\text{eff}}$ constraint. Alternatively, it may be possible to accommodate all three generations of fermions in the twin sector, provided that there is either a late-time reheating that preferentially goes into the SM sector \cite{Berezhiani:1995am, Adshead:2016xxj, asymmetricreheating, cosmomirrortwin} or a $\mathbb{Z}_2$ breaking of the Yukawa couplings \cite{Barbieri:2016zxn}. In either of the cases, the stable charged twin fermions can still interact with each other through the twin photon exchange and affect the halo formation. Further, if a component of the dark matter has acoustic oscillations, either through a twin baryon acoustic oscillation among the twin proton, twin electron, and twin photon, or through a twin lepton acoustic oscillation between the twin electron and twin neutrino through the anomaly-free gauge force U$(1)_{\hat{L}_i-\hat{L}_j}$, there are partially acoustic oscillations that can smoothly change the large-scale structure. We leave the study of this scenario to future work.

Aside from explaining the possible anomalies, analysis of structure formation in the twin sector provides additional constraints on the Twin Higgs model. Given a sizable amount of stable charged twin particles, which can be found in a large chunk of parameter space, constraints from the dark matter self-interaction enable us to set an upper bound on the twin electric coupling (Fig.~\ref{fig:SIDM}). If the charged twin particles scatter with massless twin particles, studies of the CMB and the matter power spectrum also set upper bounds on such couplings. In the coming years, the experimental precision in the values of $\Delta N_{\text{eff}}$ and $(H_0,\,\sigma_8)$ is expected to improve significantly from both the CMB and weak lensing measurements \cite{Dodelson:2016wal}. Moreover, with the progress of the N-body simulation, we anticipate to better identify the significance of the mass deficit problem from baryonic grounds. No matter whether these puzzles of the large- and small-scale structure remain significant or disappear, the coming future results will more clearly reveal the details of the Twin Higgs model, opening the door to a stronger understanding of these issues and to novel connections between the dark sector and the Hierarchy problem.

\begin{acknowledgments}

We thank Kimberly Boddy, Zackaria Chacko, Nathaniel Craig, Yanou Cui, David Curtin, Sungwoo Hong, Gustavo Marques-Tavares, Takemichi Okui,
Martin Schmaltz, Neelima Sehgal, Hai Bo Yu for helpful discussions. VP is supported by the DOE grant DE-FG02-91ER40674.
YT is 
supported in part by the National Science Foundation under grant 
PHY-1315155, and by the Maryland Center for Fundamental Physics. 
YT thanks the Aspen Center for 
Physics, which is supported by National Science Foundation grant PHY-1066293.

\end{acknowledgments}

\bibliographystyle{utphys}
\bibliography{./DMRef.bib}

\end{document}